\documentclass[aps,pra,twocolumn,superscriptaddress,longbibliography]{revtex4-2}
\usepackage{amsmath}
\usepackage{amsmath,latexsym}
\usepackage[english]{babel}
\usepackage[utf8]{inputenc}
\usepackage[colorinlistoftodos, color=green!40, prependcaption]{todonotes}
\usepackage{graphicx}
\usepackage{amsmath}
\usepackage{hyperref}
\usepackage{siunitx}
\usepackage{natbib}
\usepackage{hyperref}
\usepackage{hyperref}
 \hypersetup{ colorlinks=true, 
linkcolor=blue,
 citecolor=blue, 
 filecolor=magenta, %
 urlcolor=blue 
}

\begin{document}


\title{\textbf{Large enhancement of nonlinear optical response of graphene nanoribbon heterojunctions with multiple topological interface states} }


\author{Hanying Deng}
\email{dhy0805@alumni.sjtu.edu.cn }
\affiliation{School of Optoelectronic Engineering, Guangdong Polytechnic Normal University, Guangzhou 510665, China}
\author{Yaxin Li}
\affiliation{School of Optoelectronic Engineering, Guangdong Polytechnic Normal University, Guangzhou 510665, China}
\author{Zhihao Qu}
\affiliation{School of Optoelectronic Engineering, Guangdong Polytechnic Normal University, Guangzhou 510665, China}
\author{Jing Deng}
\affiliation{School of Optoelectronic Engineering, Guangdong Polytechnic Normal University, Guangzhou 510665, China}
\author{Yingji He}
\affiliation{School of Optoelectronic Engineering, Guangdong Polytechnic Normal University, Guangzhou 510665, China}
\author{Fangwei Ye}
\email{fangweiye@sjtu.edu.cn}
\affiliation{School of Physics and Astronomy, Shanghai Jiao Tong University, Shanghai 200240, China.}


\date{\today}
\begin{abstract} 
We investigate the nonlinear optical response of graphene nanoribbon (GNR) heterojunctions both without and with one or multiple topological interface states. By implementing a distant-neighbor quantum-mechanical (DNQM) method, we demonstrate a pronounced enhancement of the nonlinear optical response of GNR heterojunctions as the number of topological states at their interfaces increases. Specifically, we find that GNR heterojunctions with multiple topological interface states exhibit a notably stronger third-order nonlinear optical response in comparison with the similarly sized counterparts with a single topological interface state or without such states. Furthermore, we observe that the presence of topological interface states in GNR heterojunctions can induce a significant red-shift in their quantum plasmon frequency. Our results reveal the potential to enhance the nonlinear optical response at the nanoscale by increasing the number of topological interface states in graphene nanostructures or other topological systems.
\end{abstract}


\maketitle

\textit{Introduction.} Topological states have attracted widespread research attention due to their intriguing physical properties and potential applications~\cite{wen2019,kraus2012,zheng2024s}. Since the pioneering discovery of topological states in condensed matter physics, these states have been revealed in a large number of systems penetrating other field of physics, including acoustics~\cite{yang2015,ma2019t}, mechanics~\cite{huber2016} and optics~\cite{ozawa2019t,deng2016t,deng2019t,stanescu2010t}. Recently, topological states have been realized in graphene nanoribbon (GNR) heterojunctions composed of two segments of GNRs with different topological phases~\cite{zhao2021to,groning2018e}. It has been confirmed that diverse types of GNRs, including the armchair, chevron, and cove-edged GNRs, possess topological phases depending on their width, edge and end termination~\cite{cao2017topological,lee2018topological}. Furthermore, recent advances in bottom-up synthesis techniques enable atomically precise design of GNRs, including the manipulation of their edge and width~\cite{nguyen2017a,AmerChl16,Advanc17,NanoR18}, offering a plethora of GNR-based systems for further investigations and potential applications.  

With the emergence of topological states, ongoing studies have focused on the connection between topology and nonlinearity, uncovering intriguing phenomena, including the formation of topological solitons~\cite{pprlz19,nature20,sabour21}, nonlinear tuning of topological states~\cite{sabour22}, efficient harmonic generation in SSH-like transmission line~\cite{wang23}, four-wave mixing of topological edge plasmons in graphene metasurfaces~\cite{you24} and topologically enhanced nonlinear optical responses at the nanoscale~\cite{deng25}. In addition to their remarkable topological properties, GNRs inherit intense intrinsic nonlinearity from the extended graphene sheet~\cite{hendry26, wang27, cox28, cox29}, providing a natural platform for investigating the interaction between topology and nonlinearity. However, previous studies have mainly restricted to nonlinear systems with a single topological state. In this work, we extend the investigation to explore the interplay of nonlinearity and multiple topological states existing at heterojunctions composed of several segments of topologically distinct GNRs.

For graphene nanostructures with a geometrical size smaller than about 10 nm, quantum effects significantly influence their optical response, necessitating quantum mechanical calculations to describe their nonlinear optical properties~\cite{cox28, cox29, manrique30}. One commonly employed quantum mechanical method for characterizing the electronic states of nanostructured graphene is the tight-binding (TB) model~\cite{ezawa31}. However, the TB model only considers interactions between nearest-neighbor atoms, limiting its accuracy. We have recently developed a distant-neighbor quantum mechanical (DNQM) approach for calculating the optical properties of nanostructured 
graphene~\cite{deng32, clementi33,Boyd34}. This DNQM method offers greater accuracy than the traditional TB model by accounting for the interaction between the $\pi$-orbital electrons of each atom and the core potential of all atoms. In particular, our DNQM approach has been employed to computer the linear and nonlinear optical polarizabilities of nanostructured graphene with various shapes~\cite{deng25,deng32,deng35}.

In this work, we utilize the DNQM method to investigate the nonlinear optical properties of GNR heterojunctions, focusing on systems both without topological interface states and those possessing either one or multiple of such states. We demonstrate that the nonlinear optical response of GNR heterojunctions can be significantly enhanced as the number of topological interface states increases. In particular, the third-order nonlinear polarizabilities of GNR heterojunctions with multiple topological interface states are more than twice as large as those of similarly sized heterojunctions with a single topological states, and more than ten times larger than those of topologically trivial heterojunctions of similar size. Moreover, the emergence of topological states at GNR heterojunctions can shift quantum plasmon oscillations in GNR heterojunctions to a lower frequency. 

\begin{figure*}[htbp]
\centering
\includegraphics[width=0.7\textwidth]{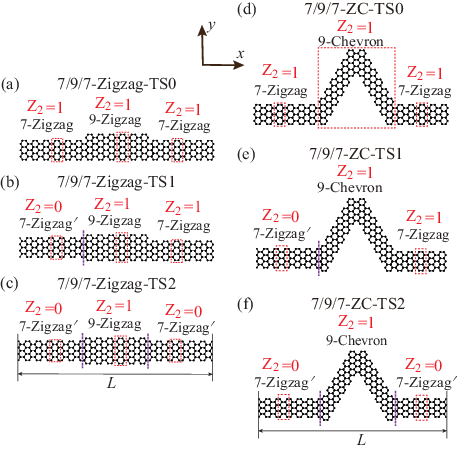}   
\caption{The geometrical configuration of GNR heterojunctions. In (a)-(c), the heterojunctions consist of three segments of AGNRs of (a) $N=7$ zigzag, $N=9$ zigzag and $N =7$ zigzag terminations, (b) $N=7$ zigzag$^{\prime}$, $N=9$ zigzag and
 $N=7$ zigzag terminations, and (c) $N=7$ zigzag$^{\prime}$, $N=9$ zigzag and $N=7$ zigzag$^{\prime}$ terminations, respectively. In (d)-(f), the heterojunctions consist of (d) two $N=7$ zigzag-terminationed AGNRs with an $N=9$ chevron GNR embedded between them, (e) an $N=7$ zigzag$^{\prime}$-terminationed AGNR, an $N=9$ chevron GNR and an $N=7$ zigzag-terminationed AGNR, and (f) two $N=7$ zigzag$^{\prime}$-terminationed AGNRs with an $N=9$ chevron GNR embedded between them, respectively. The three GNR segments in (a) and (d) are topologically equivalent, each with $Z_2=1$. The heterojunctions in (b) and (e) have one topologically nontrivial interface consisting of two GNRs with $Z_2=1$ and one GNR with $Z_2=0$, whereas those in (c) and (f) have two topologically nontrivial interfaces consisting of two GNRs with $Z_2=0$ and one GNR with $Z_2=1$. The topologically nontrivial interfaces are marked with purple dashed lines. $N$ is the number of the carbon atoms across the GNR width. The unit cell of each GNR is indicated by the red dashed rectangles.}\label{fig1}
\end{figure*}

\textit{Geometrical configurations.} The graphene nanoribbon heterojunctions considered in this work are composed of three finite GNRs, as illustrated in Fig.~\ref{fig1}. It is established that various GNRs with unit cells exhibiting spatial symmetries, such as armchair GNRs (AGNRs), chevron GNRs and cove-edged GNRs, possess symmetry-protected topological phases~\cite{cao2017topological,lee2018topological}. The topological phase of these GNRs is determined by their width, edge, and end termination, and is characterized by a $Z_2$ invariant with a value of $Z_2=1$ for topologically nontrivial ribbons and $Z_2=0$ for topologically trivial ribbons. For example, the values of $Z_2$ of $N$-AGNRs with an odd $N$, where $N$ is the number of the carbon atoms across the width of GNRs, can be expressed as ~\cite{cao2017topological}
\begin{subequations}\label{Z2odd}
\begin{align}\label{eq1}
&Z_2=\frac{1+(-1)^{[\frac{N}{3}]+[\frac{N+1}{2}]}}{2}\quad \mathrm{for zigzag-terminated},\\
\label{eq2}
&Z_2=\frac{1-(-1)^{[\frac{N}{3}]+[\frac{N+1}{2}]}}{2}\quad \mathrm{for zigzag^{\prime}-terminated},
\end{align}
\end{subequations}
respectively. Here $[x]$ is the floor function, which returns the maximum integer less than or equal to the real number  $x$. Moreover, the $N=9$ chevron GNR presented in Figs.~\ref{fig1}(d-f) was recently synthesized, and has $Z_2=1$ ~\cite{lee2018topological,NanoR18}. 

Guided by the $Z_2$ invariants of various GNRs, we construct three types of GNR heterojunctions lying in the $x-y$ plane: those without topologically nontrivial interfaces, and those with one or two topologically nontrivial interfaces, as depicted in Fig.~\ref{fig1}. The three heterojunctions shown in Figs.~\ref{fig1}(a-c) consist of the following configurations: two $N=7$ zigzag-terminated AGNRs ($Z_2=1$) with an $N=9$ zigzag-terminated AGNR ($Z_2=1$) embedded between them (Fig.~\ref{fig1}(a)); an $N=7$ zigzag$^{\prime}$-terminated ($Z_2=0$), an $N=9$ zigzag-terminated ($Z_2=1$) and an $N = 7$ zigzag-terminated ($Z_2=1$) AGNRs (Fig.~\ref{fig1}(b)); and two $N=7$ zigzag$^{\prime}$-terminated ($Z_2=0$) AGNRs with an $N=9$ zigzag-terminated AGNR ($Z_2=1$) embedded between them (Fig.~\ref{fig1}(c)). Consequently, the heterojunctions in Figs.~\ref{fig1}(a-c) have zero, one and two topologically nontrivial interfaces, as marked by the purple dashed lines, and we denote them as 7/9/7-Zigzag-TS0, 7/9/7-Zigzag-TS1 and 7/9/7-Zigzag-TS2, respectively. Note that these three heterojunctions have the same length with $L=\SI{9.23}{\nm}$, and contain the same number of carbon atoms, $N_c = 348$. We also consider three more complex heterojunctions, each composed of two $N=7$ AGNRs with either zigzag or zigzag$^{\prime}$ terminations, and an $N=9$ chevron GNR inserted between them, as shown in Figs.~\ref{fig1}(d-f). These three heterojunctions possess zero, one and two topologically nontrivial interfaces, and are correspondingly labeled as 7/9/7-ZC-TS0, 7/9/7-ZC-TS1 and 7/9/7-ZC-TS2, respectively. Moreover, the length and number of carbon atoms of these three heterojunctions in Figs.~\ref{fig1}(d-f) are identical, with $L=\SI{7.526}{\nm}$ and $N_c = 310$.

\begin{figure*}[htb]
\centering
\includegraphics[width=0.7\textwidth]{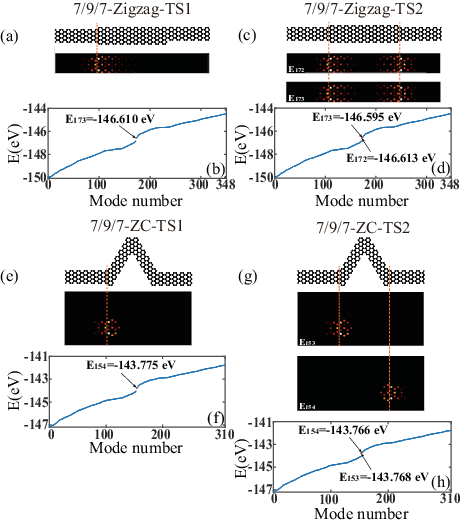}
\caption{Charge density distributions of topological interface states and energy spectra of (a, b) 7/9/7-Zigzag-TS1, (c, d) 7/9/7-Zigzag-TS2, (e, f) 7/9/7-ZC-TS1 and (g, h) 7/9/7-ZC-TS2 heterojunctions, respectively. The red dots in the energy spectra denote the topological interface states. One topological interface state exists at the topologically nontrivial interface in (a) 7/9/7-Zigzag-TS1 and (e) 7/9/7-ZC-TS1 heterojunctions. Both the 7/9/7-Zigzag-TS2 and 7/9/7-ZC-TS2 heterojunctions have two energy values corresponding to the topological interface states, as indicated by the red dots in (d) and (h). The 7/9/7-Zigzag-TS2 heterojunction supports four topological interface states with two such states having the same energy value coexisting at each of its two topologically nontrivial interfaces, whereas the 7/9/7-ZC-TS2 heterojunction possesses only two topological interface state with different energy values located at one of its topologically nontrivial interfaces. The brown dashed lines mark the topologically nontrivial interfaces.}\label{fig2}
\end{figure*}

\textit{Topological interface states and energy spectra of GNR heterojunctions.} We first calculate the charge density distributions and energy spectra of the GNR heterojunctions shown in Fig.~\ref{fig1} by using the DNQM method (see Supplemental Material for more details~\cite{SM}), the results of these calculations being summarized in Fig.~\ref{fig2}. As expected, both the 7/9/7-Zigzag-TS1 and 7/9/7-ZC-TS1 heterojunctions have one topological interface state located at their topologically nontrivial interface, as shown in Figs.~\ref{fig2}(a) and~\ref{fig2}(e), respectively. Figures ~\ref{fig2}(b) and ~\ref{fig2}(f) present the energy spectra for the 7/9/7-Zigzag-TS1 and 7/9/7-ZC-TS1 heterojunctions, respectively, with modes ordered by increasing energy. We find that the energy of the topological interface states of both the 7/9/7-Zigzag-TS1 and 7/9/7-ZC-TS1 heterojunctions resides at the HOMO-1 (where HOMO is the highest-occupied molecular orbital), with values of E=-141.61 eV and E=-143.775 eV, as denoted by the red dots in Figs. ~\ref{fig2}(b) and ~\ref{fig2}(f), respectively. 

\begin{figure*}[htb]
\centering
\includegraphics[width=0.7\textwidth]{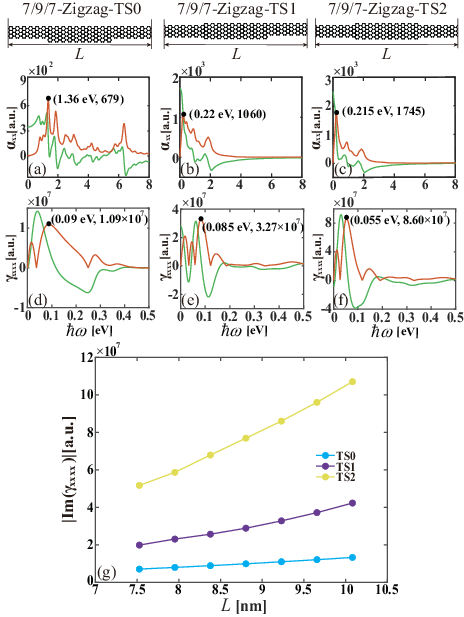}
\caption{Real and imaginary parts of linear and third-order nonlinear polarizabilities for 7/9/7-Zigzag-TS0 (a, d), 7/9/7-Zigzag-TS1 (b, e) and 7/9/7-Zigzag-TS2 (c, f) heterojunctions.  The black dots mark the resonance peaks. The red (green) curves indicate the imaginary (real) part of the polarizabilities. (g) Evolution of the peak value of THG polarizabilities on the side length of 7/9/7-Zigzag-TS0 (blue curve), 7/9/7-Zigzag-TS1 (purple curve) and 7/9/7-Zigzag-TS2 (yellow curve) heterojunctions, respectively. The calculation of polarizabilities is performed using atomic units (a.u.) with $m_{e}=e=\hbar=a_{0}=1$.}\label{fig3}
\end{figure*}

The topological interface states and energy spectrum of the 7/9/7-Zigzag-TS2 heterojunction are shown in Figs.~\ref{fig2}(c) and~\ref{fig2}(d), respectively. We find that the 7/9/7-Zigzag-TS2 heterojunction has two energy values corresponding to the topological interface states, residing at the HOMO-1 and the HOMO-2, with values being E=-146.613 eV and E=-146.595 eV, as indicated by the red dots in Fig.~\ref{fig2}(d), respectively. As presented in Fig.~\ref{fig2}(c), the 7/9/7-Zigzag-TS2 heterojunction supports four topological interface states, with two such states of the same energy coexisting at each of its two topologically nontrivial interfaces. Similarly, the 7/9/7-ZC-TS2 heterojunction also has two energy values associated with the topological interface states, having values of E=-143.768 eV and E=-143.766 eV, corresponding to the HOMO-1 and the HOMO-2, respectively, as denoted by the red dots in Fig.~\ref{fig2}(h). Note that the 7/9/7-ZC-TS2 heterojunction supports only two topological interface states with different energy values located at one of its topologically nontrivial interfaces, as shown in Fig.~\ref{fig2}(g). In contrast, no localized interface states appear at the 7/9/7-Zigzag-TS0 and 7/9/7-ZC-TS0 heterojunctions, as they do not possess topologically nontrivial interfaces. 

\textit{Linear and nonlinear optical response of GNR heterojunctions.} We next study the linear and nonlinear optical response of GNR heterojunctions shown in Fig.~\ref{fig1}. Due to their structural symmetry, the second-order nonlinear process is not allowed. Thus, we focus on the linear and third-order optical response, assuming that the incident electric field is $x$-polarized. We first consider the 7/9/7-Zigzag-TS0, 7/9/7-Zigzag-TS1 and 7/9/7-Zigzag-TS2 heterojunctions. As discussed above, these three heterojunctions consist of the same number of carbon atoms and have the same side length. Moreover, the 7/9/7-Zigzag-TS0 does not support topological interface states, whereas the 7/9/7-Zigzag-TS1 and 7/9/7-Zigzag-TS2 heterojunctions possess one and four topological interface states, respectively. The results of the calculations for the linear and third-order nonlinear polarizabilities of 7/9/7-Zigzag-TS0, 7/9/7-Zigzag-TS1 and 7/9/7-Zigzag-TS2 heterojunctions are summarized in Fig.~\ref{fig3}(see Supplemental Material for more details~\cite{SM}).

Figures~\ref{fig3}(a-c) present the frequency dependence of real and imaginary parts of linear polarizabilities, $\alpha_{xx}(\omega)$, of 7/9/7-Zigzag-TS0, 7/9/7-Zigzag-TS1 and 7/9/7-Zigzag-TS2 heterojunctions, respectively. It can be seen that the frequency of the most pronounced resonance peaks in the linear polarizability spectra of the topologically nontrivial heterojunctions, namely, the 7/9/7-Zigzag-TS1 and 7/9/7-Zigzag-TS2 heterojunctions, is significantly red-shifted compared to that of the topologically trivial 7/9/7-Zigzag-TS0 heterojunction. To be more specific, the most pronounced resonance peak of the 7/9/7-Zigzag-TS0 heterojunction is at  \SI{1.36}{\eV}, whereas the 7/9/7-Zigzag-TS1 and 7/9/7-Zigzag-TS2 heterojunctions have their most pronounced resonance peaks located at \SI{0.22}{\eV} and \SI{0.215}{\eV}, as denoted by the black dots in Figs.~\ref{fig3}(a-c), respectively. 

\begin{figure*}[htb]
\centering
\includegraphics[width=0.7\textwidth]{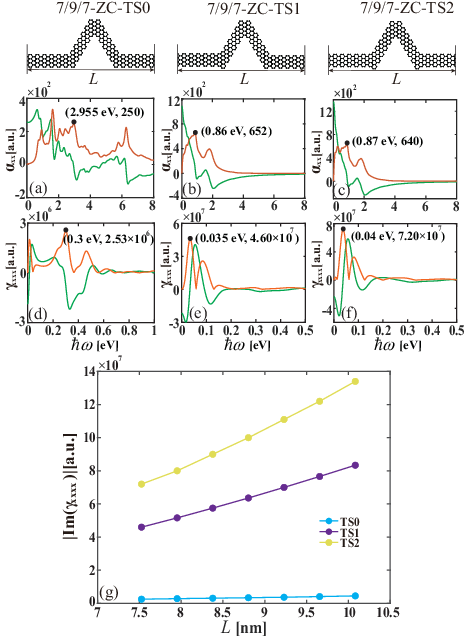}
\caption{Real and imaginary parts of linear and third-order nonlinear polarizabilities of 7/9/7-ZC-TS0 (a, d), 7/9/7-ZC-TS1 (b, e) and 7/9/7-ZC-TS2 (c, f) heterojunctions. The black dots mark the resonance peaks. The red (green) curves indicate the imaginary (real) part of the polarizabilities. (g) Evolution of the peak value of THG polarizabilities on the side length of 7/9/7-ZC-TS0 (blue curve), 7/9/7-ZC-TS1 (purple curve) and 7/9/7-ZC-TS2 (yellow curve) heterojunctions, respectively.}\label{fig4}
\end{figure*}

In Figs.~\ref{fig3}(d-f), we show the real and imaginary parts of the third-order nonlinear polarizabilities, $\gamma$$_{xxxx}(3\omega)$, corresponding to the third-harmonic generation (THG), of 7/9/7-Zigzag-TS0, 7/9/7-Zigzag-TS1 and 7/9/7-Zigzag-TS2 heterojunctions, respectively. The spectra clearly indicate that the peak of the imaginary part of the THG polarizability coincides with the zero point of the real part, defining the existence of quantum plasmons~\cite{manrique30,deng32}. Comparing the three THG polarizability spectra presented in Figs.~\ref{fig3}(d-f), we find that the THG polarizability of 7/9/7-Zigzag-TS2 heterojunction with multiple topological states is more than twice as large as that of similarly sized 7/9/7-Zigzag-TS1 heterojunction with a single topological state, and more than seven times larger than that of topologically trivial 7/9/7-Zigzag-TS0 heterojunction of similar size. Moreover, similar to the linear case, the resonance frequency of the nonlinear quantum plasmons of both the 7/9/7-Zigzag-TS1 and 7/9/7-Zigzag-TS2 heterojunctions is red-shifted as compared with that of the 7/9/7-Zigzag-TS0 heterojunction. These remarkable results are a consequence of the topological interface states participate in the transitions. More specifically, as labeled by the black dot in Fig.~\ref{fig3}(e), the most pronounced resonance peak in the spectrum of THG polarizabilily of the 7/9/7-Zigzag-TS1 heterojunction is located at ${\hbar}\omega=\SI{0.085}{\eV}$, which is associated with the transition from the HOMO-1 (which corresponds to the topological states) to the LUMO+1. In the 7/9/7-Zigzag-TS2 heterojunction, the most pronounced resonance peak located at ${\hbar}\omega=\SI{0.055}{\eV}$, indicated by the red dot in  Fig.~\ref{fig3}(f), has contributions from the transitions from the HOMO-1 to the LUMO+1 and from the HOMO-2 to the LUMO, with both the HOMO-1 and HOMO-2 corresponding to the topological interface states. Therefore, the 7/9/7-Zigzag-TS2 heterojunction, possessing more topological interface states involved in these transitions, exhibits the THG more than twice as large as that of 7/9/7-Zigzag-TS1 heterojunction with only one topological interface state.

We show in Fig.~\ref{fig3}(g) the evolution of the peak value of THG polarizabilities on the side length of 7/9/7-Zigzag-TS0, 7/9/7-Zigzag-TS1 and 7/9/7-Zigzag-TS2 heterojunctions, respectively. This figure reveals several important conclusions. First, the more topological interface states a GNR heterojunction possesses, the larger its THG. Second, as expected, the THG polarizabilities of all three heterojunctions increase with their side length. Third, the rate of the increase of THG polarizabilities with side length also rises with the number of the topological interface states present in the heterojunctions. Notably, the 7/9/7-Zigzag-TS2 heterojunction, which possesses four topological interface states, exhibits the highest rate of increase of THG polarizabilities with its side length.

We now move on to the linear and third-order nonlinear optical response of the 7/9/7-ZC-TS0, 7/9/7-ZC-TS1 and 7/9/7-ZC-TS2 heterojunctions. As mentioned above, the 7/9/7-ZC-TS1 and 7/9/7-ZC-TS2 heterojunctions have one and two topological interface states, respectively, while no topological interface states exist at 7/9/7-ZC-TS0 heterojunction. In Figs.~\ref{fig4}(a-c), we present the real and imaginary parts of the linear polarizabilities,  $\alpha_{xx}(\omega)$, of 7/9/7-ZC-TS0, 7/9/7-ZC-TS1 and 7/9/7-ZC-TS2 heterojunctions, respectively. Comparing these spectra, we can see that again the emergence of the topological interface states can lead to a significant red-shift in the resonance frequency of quantum plasmons. Particularly, the topologically nontrivial 7/9/7-ZC-TS2 and 7/9/7-ZC-TS1 heterojunctions have the frequency of quantum plasmons located at 0.87 eV and 0.86 eV, whereas the quantum plasmon frequency of 7/9/7-ZC-TS0 heterojunction is at 2.955 eV, as labeled by the black dots in Figs.~\ref{fig4}(a-c), respectively. 

\begin{figure*}[htb]
\centering
\includegraphics[width=0.7\textwidth]{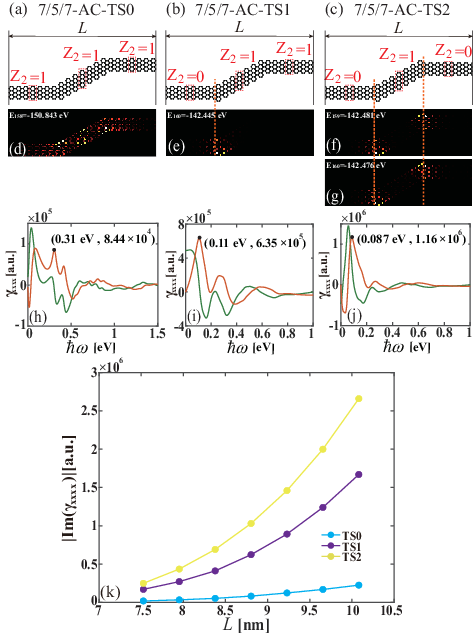}
\caption{GNR heterojunctions consist of (a) two $N=7$ zigzag-terminated AGNRs with a $K=5$ cove-edged GNR inserted between them, (b) an $N=7$ zigzag$^{\prime}$-terminated and an $N=7$ zigzag-terminated AGNRs with a $K=5$ cove-edged GNR inserted between them, and (c) two $N=7$ zigzag$^{\prime}$-terminated AGNRs with a $K=5$ cove-edged GNR inserted between them. The unit cell of each GNR forming the heterojunctions is indicated by the red dashed rectangles. The three heterojunctions in (a)-(c) possess zero, one and two topologically nontrivial interfaces, respectively. The brown dashed lines mark the topologically nontrivial interfaces. Charge density distributions of (d) 7/5/7-AC-TS0, (e) 7/5/7-AC-TS1 and (f, g) 7/5/7-AC-TS2 heterojunctions. Real and imaginary parts of THG polarizabilities of (h) 7/5/7-AC-TS0, (i) 7/5/7-AC-TS1 and (j) 7/5/7-AC-TS2 heterojunctions. The red (green) curves indicate the imaginary (real) part of the polarizabilities. (k) Evolution of the peak value of THG polarizabilities on the side length of 7/5/7-AC-TS0 (blue curve), 7/5/7-AC-TS1 (purple curve) and 7/5/7-AC-TS2 (yellow curve) heterojunctions, respectively.}\label{fig5}
\end{figure*}

The THG polarizabilities of 7/9/7-ZC-TS0, 7/9/7-ZC-TS1 and 7/9/7-ZC-TS2 heterojunctions, are shown in Figs.~\ref{fig4}(d)-4(f). Similar to what we observed in Fig.~\ref{fig3}, the magnitude of THG polarizabilities of these heterojunctions depends significantly on the number of topological interface states that they possess. In particular, due to having two topological interface states, the 7/9/7-ZC-TS2 heterojunction has the largest THG polarizabilities as compared to the similar sized 7/9/7-ZC-TS1 heterojunction with one topological interface state and the 7/9/7-ZC-TS0 heterojunction without topological interface states, as denoted by the black dots in Figs.~\ref{fig4}(d-f). Moreover, the appearance of topological interface states also results in a remarkable red-shift of the resonance frequency of the third-order nonlinear quantum plasmons of the 7/9/7-ZC-TS1 and 7/9/7-ZC-TS2 heterojunctions, as compared with that of 7/9/7-ZC-TS0 heterojunction. Figure~\ref{fig4}(g) displays an overview of the dependence of the maximum THG polarizabilities on the side length of 7/9/7-ZC-TS0, 7/9/7-ZC-TS1 and 7/9/7-ZC-TS2 heterojunctions. Again, it can be observed that both the magnitude and the rate of increase in length of the THG polarizabilities increase with the number of topological interface states in the heterojunctions. Additionally, the magnitude of THG polarizabilities of these three heterojunctions increases with their side length.

Finally, we consider three GNR heterojunctions composed of two $N=7$ AGNRs of either zigzag or zigzag$^{\prime}$ terminations, and a $K=5$ cove-edged GNR with a tilting angle of \ang{30}, as illustrated in Figs.~\ref{fig5}(a-c). Here, $K$ is the number of zigzag chains forming the width of the cove-edged GNR. The topological properties of cove-edged GNRs are also determined by their width, edge and end termination~\cite{lee2018topological}. The $Z_2$ invariants of GNRs forming the heterojunctions are labeled in Figs.~\ref{fig5}(a-c). These three GNR heterojunctions possess zero, one and two topologically nontrivial interfaces, and thus are denoted as 7/5/7-AC-TS0, 7/5/7-AC-TS1 and 7/5/7-AC-TS2, respectively. Moreover, they have the same length of $L=\SI{8.8}{\nm}$ and contain the same number of carbon atoms of $N_c=322$. As presented in Figs.~\ref{fig5}(e-g), the 7/5/7-AC-TS1 heterojunction supports one topological interface state located at its topologically nontrivial interface, while the 7/5/7-AC-TS2 heterojunction possesses four topological interface states with two such states of the same energy coexisting at each of its two topologically nontrivial interfaces. In contrast, no located interface states exist in 7/5/7-AC-TS0 heterojunction, and its charge profile is extended throughout the heterojunction, as shown in Fig.~\ref{fig5}(d).

Using the DNQM method, we have computed the THG polarizabilities of the 7/5/7-AC-TS0, 7/5/7-AC-TS1 and 7/5/7-AC-TS2 heterojunctions, with results shown in Figs.~\ref{fig5}(h-j). Comparing these spectral, we can draw conclusions that are consistent with those from Fig.~\ref{fig3} and Fig.~\ref{fig4}. More specifically, the presence of topological interface states leads to a significant red-shift of the nonlinear quantum plasmon frequency. Moreover, the magnitude of the THG polarizabilities of GNR heterojunctions increases with the increase of the number of topological interface states at the heterojunctions. This feather can be further demonstrated by the dependence of the peak value of THG polarizabilities on the side length of 7/5/7-AC-TS0, 7/5/7-AC-TS1 and 7/5/7-AC-TS2 heterojunctions, as shown in Fig.~\ref{fig5}(k). It can be seen from this figure that the 7/5/7-AC-TS2 heterojunction with four topological interface states has the largest THG polarizability, as compared with the similar sized 7/5/7-AC-TS0 and 7/5/7-AC-TS1 heterojunctions.

\textit{Conclusion.} We have employed the DNQM approach to investigate the nonlinear optical response of GNR heterojunctions both without and with one or multiple topological interface states. Our analysis has revealed that the nonlinear optical response of GNR heterojunctions can be significantly enhanced by increasing the number of topological interface states at the heterojunctions. In particular, the third-order nonlinear polarizabilities of GNR heterojunctions with multiple topological states are more than twice as large as those of similarly sized heterojunctions with a single topological states, and over ten times larger than those of topologically trivial heterojunctions of similar size, despite having the same side length and contain the same number of carbon atoms. Additionally, the frequency of quantum plasmons of GNR heterojunctions with topological interface states is significantly red-shifted compared with that of GNR heterojunctions without such states. Our work offers a promising method to enhance the nonlinear optical response at the nanoscale, thus holding strong potential for nonlinear nanophotonics and quantum optics.

\textit{Acknowledgments.} This work was supported by the National Natural Science Foundation of China (Grant Nos. 12104104, 62175042), the Natural Science Foundation of Guangdong Province (Grant No. 2019A1515011499), the Start-up Funding of Guangdong Polytechnic Normal University (Grant No. 2021SDKYA165) and the Guangdong Department of Education Projects of Improving Scientific Research Capabilities of Key Subjects Construction (Grant No. 2022ZDJS016).


\begin{thebibliography}{99}
\bibitem{wen2019}
X.G. Wen, Choreographed entanglement dances: Topological states of quantum matter, Science \textbf{363}, eaal3099
(2019).
\bibitem{kraus2012}
Y. E. Kraus, Y. Lahini, Z. Ringel, M. Verbin, and O. Zilberberg, Topological states and adiabatic pumping in
quasicrystals, Phys. Rev. Lett. \textbf{109}, 106402 (2012).
\bibitem{zheng2024s}
C. Zheng and X. Liu, Superconductivity and topological quantum states in two-dimensional moir´e superlattices,
Quantum Front. \textbf{3}, 17 (2024).
\bibitem{yang2015}
Z. Yang, F. Gao, X. Shi, X. Lin, Z. Gao, Y. Chong, and B. Zhang, Topological acoustics, Phys. Rev. Lett. \textbf{114}, 114301 (2015).
\bibitem{ma2019t}
G. Ma, M. Xiao, and C. T. Chan, Topological phases in acoustic and mechanical systems, Nat. Rev. Phys. \textbf{1}, 281 (2019).
\bibitem{huber2016}
S. D. Huber, Topological mechanics, Nat. Phys. \textbf{12}, 621 (2016).
\bibitem{ozawa2019t}
T. Ozawa, H. M. Price, A. Amo, N. Goldman, M. Hafezi, L. Lu, M. C. Rechtsman, D. Schuster, J. Simon, O. Zilberberg, et al., Topological photonics, Rev. Mod. Phys. \text{91}, 015006 (2019)
\bibitem{deng2016t}
H. Deng, X. Chen, N. C. Panoiu, and F. Ye, Topological surface plasmons in superlattices with changing sign of
the average permittivity, Opt. Lett. \textbf{41}, 4281 (2016).
\bibitem{deng2019t}
H. Deng, Y. Chen, C. Huang, and F. Ye, Topological interface modes in photonic superlattices containing
negative-index materials, Europhys. Lett. \textbf{124}, 64001 (2019).
\bibitem{stanescu2010t}
T. D. Stanescu, V. Galitski, and S. Das Sarma, Topological states in two-dimensional optical lattices, Phys. Rev. A \textbf{82}, 013608 (2010).


\bibitem{zhao2021to}
F. Zhao, T. Cao, and S. G. Louie, Topological phases in
graphene nanoribbons tuned by electric fields, Phys. Rev. lett. \textbf{127}, 166401 (2021).
\bibitem{groning2018e}
O. Gr \"{o}ning, S. Wang, X. Yao, C. A. Pignedoli, G. Borin Barin, C. Daniels, A. Cupo, V. Meunier,
X. Feng, A. Narita, et al., Engineering of robust topological quantum phases in graphene nanoribbons, Nature
\textbf{560}, 209 (2018).
\bibitem{cao2017topological}
T. Cao, F. Zhao, and S. G. Louie, Topological phases in graphene nanoribbons: junction states, spin centers, and quantum spin chains, Phys. Rev. Lett. \textbf{119}, 076401 (2017).
\bibitem{lee2018topological}
 Y.-L. Lee, F. Zhao, T. Cao, J. Ihm, and S. G. Louie, Topological phases in cove-edged and chevron graphene nanoribbons: Geometric structures, $Z_2$ invariants, and
junction states, Nano Lett. \textbf{18}, 7247 (2018).
\bibitem{nguyen2017a}
G. D. Nguyen, H.-Z. Tsai, A. A. Omrani, T. Marangoni,M. Wu, D. J. Rizzo, G. F. Rodgers, R. R. Cloke,R. A. Durr, Y. Sakai, et al., Atomically precise graphene
nanoribbon heterojunctions from a single molecular precursor, Nat. Nanotechnol. \textbf{12}, 1077 (2017).


\bibitem{AmerChl16}
Z. Liu, Z. Chen, C. Wang, et al., Bottom-up, on-surface-synthesized armchair graphene nanoribbons for ultra-high-power micro-supercapacitors,  J. Am. Chem. Soc. \textbf{142}, 17881 (2020).
\bibitem{Advanc17}
 Z. Chen, A. Narita, K. M{\"u}llen, Graphene nanoribbons: on-surface synthesis and integration into electronic devices. Adv. Mater. \textbf{32}, 2001893 (2020).
\bibitem{NanoR18}
 X. Liu, G. Li, A. Lipatov, T. Sun and A. Sinitskii, Chevron-type graphene nanoribbons with a reduced energy band gap: Solution synthesis, scanning tunneling microscopy and electrical characterization, Nano Res.\textbf{13} (2020).



\bibitem{pprlz19}
W. Zhang, X. Chen, Y. V. Kartashov, V. V. Konotop, and F. Ye, Coupling of edge states and topological Bragg solitons, Phys. Rev. Lett. \textbf{123},  254103 (2019).

\bibitem{nature20}
J. Veenstra, O. Gamayun, X. Guo, A. Sarvi, C. V. Meinersen, and C. Coulais, Non-reciprocal topological solitons
in active metamaterials, Nature \textbf{627}, 528 (2024).

\bibitem{sabour21}
K. Sabour and Y. V. Kartashov, Topological solitons in coupled su–schrieffer–heeger waveguide arrays, Opt.
Lett. \textbf{49}, 3580 (2024).

\bibitem{sabour22}
S. Xia, D. Kaltsas, D. Song, I. Komis, J. Xu, A. Szameit, H. Buljan, K. G. Makris, and Z. Chen, Nonlinear tuning of pt symmetry and non-hermitian topological states,
Science \textbf{372}, 72 (2021).
\bibitem{wang23}
Y. Wang, L.-J. Lang, C. H. Lee, B. Zhang, and Y. Chong,
Topologically enhanced harmonic generation in a nonlinear transmission line metamaterial, Nat. Commun. \textbf{10}, 1102 (2019).

\bibitem{you24}
 J. W. You, Z. Lan, and N. C. Panoiu, Four-wave mixing of topological edge plasmons in graphene metasurfaces,
Sci. Adv. \textbf{6}, eaaz3910 (2020).

\bibitem{deng25}
 H. Deng, Z. Qu, Y. He, C. Huang, N. C. Panoiu, and F. Ye, Topologically enhanced nonlinear optical response
of graphene nanoribbon heterojunctions, Quantum Front. \textbf{2}, 11 (2023).

\bibitem{hendry26}
E. Hendry, P. J. Hale, J. Moger, A. Savchenko, and
S. A. Mikhailov, Coherent nonlinear optical response of
graphene, Phys. Rev. Lett. \textbf{105}, 097401 (2010)

\bibitem{wang27}
Y. Wang, M. Tokman, and A. Belyanin, Second-order
nonlinear optical response of graphene, Phys. Rev.
B \textbf{94}, 195442 (2016).
\bibitem{cox28}
 J. D. Cox, I. Silveiro, and F. J. Garc´ıa de Abajo, Quantum effects in the nonlinear response of graphene plasmons, ACS Nano \textbf{10}, 1995 (2016).
\bibitem{cox29}
J. D. Cox and F. J. Garcia de Abajo, Nonlinear graphene
nanoplasmonics, Acc. Chem. Res. \textbf{52}, 2536 (2019).
\bibitem{manrique30}
D. Z. Manrique, J. W. You, H. Deng, F. Ye, and N. C.
Panoiu, Quantum plasmon engineering with interacting
graphene nanoflakes, J. Phys. Chem. C \textbf{121}, 27597 (2017).

\bibitem{ezawa31}
M. Ezawa, Metallic graphene nanodisks: Electronic and magnetic properties, Phys. Rev. B \textbf{76}, 245415 (2007).

\bibitem{deng32}
H. Deng, D. Z. Manrique, X. Chen, N. C. Panoiu, and F. Ye, Quantum mechanical analysis of nonlinear optical
response of interacting graphene nanoflakes, APL Photon. \textbf{3} (2018)

\bibitem{clementi33}
E. Clementi and D.-L. Raimondi, Atomic screening constants from scf functions, J. Chem.
Phys. \textbf{38}, 2686 (1963).

\bibitem{Boyd34}
R. W. Boyd, Nonlinear optics 3rd ed. Taylor $\&$ Francis, London (2003).

\bibitem{deng35}
H. Deng, C. Huang, Y. He, and F. Ye, Quantum plasmon enhanced nonlinear wave mixing in graphene nanoflakes,
Chin. Phys. B \textbf{30}, 044213 (2021).

\bibitem{SM}
See Supplemental Material for (1) calculation of charge density distributions and eigenenergy spectra of GNR heterojunctions; (2)  calculation of the linear and third-order nonlinear optical response of GNR heterojunctions.


\end{thebibliography}
\end{document}